\documentclass[aps,twocolumn,showpacs,preprintnumbers,floatfix]{revtex4}
\usepackage{graphicx}% Include figure files
\usepackage{epsfig}
\usepackage{dcolumn}% Align table columns on decimal point
\usepackage{amsmath}

\begin{document}

\title{Exotic vector charmonium and its leptonic decay width}

\author{\small
Ying Chen,${}^{1,2}$\footnote{cheny@ihep.ac.cn} Wei-Feng Chiu,${}^{1}$ Ming Gong,${}^{1,2}$
Long-Cheng Gui,${}^{3,4}$ and Zhaofeng Liu${}^{1,2}$ }

\affiliation{%
$^1$ Institute of High Energy Physics, Chinese Academy of Sciences, Beijing 100049, China\\
$^2$ Theoretical Center for Science Facilities, Chinese Academy of Sciences, Beijing 100049, China\\
$^3$ Department of Physics and Synergetic Innovation Center for Quantum Effects and Applications,\\
          Hunan Normal University, Changsha 410081, China \\
$^4$ Key Laboratory of Low-Dimensional Quantum Structures and Quantum Control of \\
Ministry of Education, Changsha 410081, China }

\begin{abstract}
We propose a novel type of interpolating field operators, which manifests the hybrid-like
configuration that the charm quark-antiquark pair recoils against gluonic degrees of freedom. A
heavy vector charmonium-like state with a mass of $4.33(2)\,{\rm GeV}$ is disentangled from the
conventional charmonium states in the quenched approximation. This state has affinity for the
hybrid-like operators but couples less to the relevant quark bilinear operator. We also try to
extract its leptonic decay constant and give a tentative upper limit that it is less than one tenth
of that of $J/\psi$, which corresponds to a leptonic decay width about dozens of eV.  The
connection of this state with $X(4260)$ is also discussed.
\end{abstract}

\pacs{12.38.Gc, 13.30.Ce, 14.40.Rt}\maketitle

\section{Introduction}

$X(4260)$ was observed by many experiments as a $\pi\pi J/\psi$ resonance structure in the initial
state radiation (ISR) process $e^+e^-\rightarrow \gamma_{\rm ISR}J/\psi\pi\pi$~\cite{Aubert:2005rm,
Coan:2006rv,Yuan:2007sj}. Its resonance parameters are determined now to be $M_X=4251(9)$ MeV and
$\Gamma_X=120(12)$ MeV~\cite{Agashe:2014kda}. According to its production mode, $X(4260)$ must have
the quantum number $J^{PC}=1^{--}$. In addition, the ratio of $X(4260)\rightarrow \pi^+\pi^-
J/\psi$ and $X(4260)\rightarrow \pi^0\pi^0 J/\psi$ events observed by the CLEO collaboration is
consistent with $X(4260)$ being an isoscalar. In other words, $X(4260)$ has the same quantum number
as that of vector charmonia $J/\psi$, $\psi'$, etc. However, in contrast to $\psi$ states,
$X(4260)$ has not been observed directly in the $e^+e^-$ annihilation yet. On the other hand, its
mass is well above the $D\bar{D}$ threshold, but it has been observed only in the $J/\psi
\pi^+\pi^-$ system instead of $D\bar{D}$ ones. These facts may imply that $X(4260)$ has a large
branch fraction for the $J/\psi\pi^+\pi^-$ decay mode. Thus, the small combined width
$\Gamma(X\rightarrow e^+e^-)Br(X\rightarrow J/\psi\pi\pi)=9.2\pm 1.0\,{\rm eV}$ can be understood
as that $X(4260)$ has a very small $e^+e^-$ width. These features motivate conjectures that
$X(4260)$ might be an exotic state, for example, a hybrid charmonium~\cite{Zhu:2005hp,
Close:2005iz, Kou:2005gt}. Anyway, more theoretical information for $X(4260)$ is needed in order to
unravel its nature, among which the leptonic decay width of $X(4260)$ is an important quantity.

As far as the hybrid charmonium is concerned, extensive lattice QCD studies have been devoted to
$J^{PC}=1^{-+}$ channel. In the constituent quark model picture, this quantum number cannot appear
in the $q\bar{q}$ system, therefore it is usually conjectured that additional degrees of freedom
should be involved and the minimum configuration can be $q\bar{q}g$ where $g$ is a constituent
gluon. The corresponding $q\bar{q}g$ interpolation fields are used in lattice calculations which
predict the mass of the $1^{-+}$ charmonum-like state to be around $4.3$
GeV~\cite{Lacock:1996vy,Bernard:1997ib,Liao:2002rj,Bernard:2003jd,Mei:2002ip,Dudek:2009qf,Liu:2012ze,Yang:2012gz,Dudek:2008sz,Dudek:2009kk}.
Similar studies have been also extended to investigate possible hybrids with the conventional
quantum number, but the challenging task is to distinguish these states from conventional mesons.
The state-of-art approach for this goal is the variational method based on large enough operator
sets built through sophisticated methods. In the vector channel, there is a state observed with a
mass round $4.4\,{\rm GeV}$~\cite{Liu:2012ze}, which couples weakly to the quark bilinear operator
but seems intimate with quark-antiquark-gluon operators.

Generally speaking, the appearance of an interpolating field operator does not necessarily reflect
the inner structure of a hadron state. However, for heavy quark systems where the non-relativistic
picture may be available to some extent, the coupling of the operator to a specific state may bear
some useful information of its status. Taking a mesonic hybrid for example, even though it is an
ambiguous concept from the point of quantum chromodynamics (QCD) , it is always thought of a hadron
state made up of a quark-antiquark pair plus a constituent gluonic component in the constituent
quark model picture. Of course one can also relax the definition of a hybrid to an exotic object
which has additional degree of freedom apart from the constituent quarks. This kind of additional
degree of freedom can be a fluctuating flux tube in the flux tube model, the color bag of the MIT
bag model, etc. Anyway, the essence of the exotic nature of a meson state is that the constituent
$q\bar{q}$ pair acquires a center-of-mass motion by recoiling against the additional degrees of
freedom, which is distinct from the conventional hadron states. This is our starting point to build
a novel type of hybrid-like operators. We split the charm quark-antiquark pair component and the
gluon field in $q\bar{q}g$ operator into two parts with different spatial separations. In the
momentum space, this manifests the center-of-mass motion of $q\bar{q}$ pair in the rest frame of
the state. We calculate the correlation functions of these operators, from which we try to extract
the possible exotic charmonium state. Since the operators with different spatial separation provide
different correlation functions, we fit them simultaneously along with the correlation function
involving the electromagnetic current to obtain the decay constants of the states which contribute
significantly.

This paper is organized as follows: Section 2 contains the description of the construction of the
new lattice interpolation operators for the hybrid-like vector meson. The lattice parameters and
the numerical techniques are presented in Sec. 3. We discuss our results and their connections with
$X(4260)$ in Sec. 4. The conclusions and a summary can be found in Sec. 5.

\section{New interpolation field for exotic vector charmonium}

In this work, we will focus on the exotic vector charmonium ($J^{PC}=1^{--}$) by assuming a
hybrid-like configuration $c\bar{c}g$. A simple and straightforward local operator possibly
reflecting this constituent configuration is $O_i^{(H)}(x)=\bar{c}^a(x)\gamma_5 c^b(x)B_i^{ab}(x)$,
where $a,b$ are color indices, $i$ the spatial index, and
$B_i^{ab}(x)=\frac{1}{2}\epsilon_{ijk}F_{jk}^{ab}$ the chromomagnetic field tensor. This kind of
operator can be compared with the commonly used quark bilinear operator for the vector
$O_i^{(M)}=\bar{c}\gamma_i c (x)$. In order to find the nonrelativistic form of these interpolation
operators, we use the Foldy-Wouthuysen-Tani transformation~\cite{Foldy:1950} to decompose the charm
quark and antiquark fields (Dirac spinor) in terms of the Pauli spinors $\phi/\phi^\dagger$ which
annihilates/creates a charm quark, and $\chi/\chi^\dagger$ which creates/annihilates a charm
antiquark. The explicit expressions of the operators $O_i^{(H)}(x)$ and $O_i^{(M)}$ to the lowest
order of the nonrelativistic approximation can be written as
\begin{eqnarray}
O_i^{(H)}&\equiv& \bar{c}^a\gamma_5 c^b B_i^{ab}\rightarrow
\chi^{a\dagger}\phi^b B_i^{ab}+O(\frac{1}{m_c}),\nonumber\\
O_i^{(M)}&\equiv& \bar{c}^a\gamma_i c^a\rightarrow \chi^{a\dagger}\sigma_i\phi^a+O(\frac{1}{m_c}),
\end{eqnarray}
where one can see that the block $\chi^{a\dagger}\phi^b$ of the $O_i^{(H)}$ operator is a spin
singlet and color octet, while that of $O_i^{(M)}$ is a spin triplet and color singlet. Intuitively
$O_i^{(H)}$ couples more to a state of spin singlet charm quark-antiquark pair and less to a state
of spin triplet $\bar{c}c$ component owing to the heavy quark mass suppression for the spin
flipping of a heavy quark, and vice versa for $O_i^{(M)}$. In order to resemble the center-of-mass
motion of the $c\bar{c}$ recoiling against an additional degree of freedom, we split the operator
$O_i^{(H)}$ into two spatial parts, $\bar{c}^a\gamma_5 c^b$ and $B_{i}^{ab}$, separated by an
explicit spatial displacement $\mathbf{r}$. In a fixed gauge (the Coulomb gauge in this work), we
get a set of spatially extended operators,
\begin{equation}
O_i^{(H)}(\mathbf{x},t;\mathbf{r})=(\bar{c}^a\gamma_5 c^b)(\mathbf{x},t)B_i^{ab}(\mathbf{x+r},t).
\end{equation}
It is expected that the coupling of this type of operator to the conventional charmonia (without
the center-of-mass motion of charm quark-antiquark pair in the non-relativistic picture) would be
suppressed, while the coupling to the exotic state can be enhanced.

\section{Numerical details}
We use the tadpole-improved gauge action~\cite{Morningstar:1997ff,Morningstar:1999rf,Chen:2005mg}
to generate gauge configurations on anisotropic lattices with the temporal lattice spacing much
finer than the spatial one. The aspect ratio takes $\xi=a_s/a_t=5$, where $a_s$ and $a_t$ are the
spatial and temporal lattice spacing, respectively. Two lattices $L^3\times T=8^3\times
96(\beta=2.4)$ and $12^3\times 144(\beta=2.8)$ with different lattice spacings are used to check
the discretization artifacts and the relevant input parameters are listed in
Table~\ref{tab:lattice}, where $a_s$ values are determined from $r_0^{-1}=410(20)$ MeV. We use the
tadpole-improved clover action to calculate the quark propagators. The relevant parameters in the
fermion action are tuned carefully by requiring that the physical dispersion relations of vector
and pseudoscalar mesons are correctly reproduced at each bare quark mass~\cite{Liu:2001ss,
Liu:2005tc}. The bare charm quark masses at different $\beta$ are determined from the physical mass
of $J/\psi$, $m_{J/\psi}=3.097$ GeV. The spatial extension of both lattices is $\sim 1.7\,{\rm
fm}$, which is tested to be large enough for charmonium states. The ground state masses of $1S$ and
$1P$ charmonia calculated on these two lattices (see Fig.~2 and Table II of
Ref.~\cite{Yang:2012mya} for the details) show that the finite $a_s$ effects are small. Since the
spatial extended interpolation operators $O_i^{(H)}$ discussed above are gauge variant, we carry
out the calculation of the quark propagators and correlation functions after transforming each
configuration to the Coulomb gauge.

\begin{table}[t]
\centering\caption{\label{tab:lattice}The input parameters for the calculation. Values of the
coupling $\beta$, anisotropy $\xi$, the lattice size, and the number of measurements are listed.
$a_s/r_0$ is determined by the static potential, the first error of $a_s$ is the statistical error
and the second one comes from the uncertainty of the scale parameter $r_0^{-1}=410(20)$ MeV.}
\begin{ruledtabular}
\begin{tabular}{cccccc}
     $\beta$ &  $\xi$  &  $a_s$  & $La_s$(fm)& $L^3\times T$ & $N_{\rm conf}$ \\\hline
       2.4  & 5  &  0.222(2)(11) & $\sim 1.78$ &$8^3\times 96$ & 1000 \\
      2.8  & 5  &  0.138(1)(7) & $\sim 1.66 $&$12^3\times 144$ & 1000  \\
\end{tabular}
\end{ruledtabular}
\end{table}

\subsection{Data analysis strategy}
Our first task is to verify the existence of the exotic vector charmonium. We use the following
source operator to calculate the correlation functions,
\begin{equation}
O_i^{(W)}(\tau)=\sum\limits_{\mathbf{y,z}}\bar{c}^a(\mathbf{y},\tau)\gamma_5
B_i^{ab}(\mathbf{z},\tau) c^b(\mathbf{z},\tau),
\end{equation}
where $\tau$ refers to the source time slice. For the sink operator $O_i^{(H)}$, the two-point
functions we calculate are
\begin{eqnarray}\label{two-point}
C^{(H)}(\mathbf{r},t;\tau)&=&\frac{1}{3}\sum\limits_{\mathbf{x},i}\langle
0|O_i^{(H)}(\mathbf{x},t;\mathbf{r})O_i^{(W)\dagger}(\tau)|0\rangle\nonumber\\
&=&\frac{1}{3}\sum\limits_{\mathbf{x,y,z},i} Tr \left\langle
S_F^\dagger(\mathbf{x},t;\mathbf{y},\tau)
B_i(\mathbf{x+r},t)\right.\nonumber\\
&&\left. \times S_F (\mathbf{x},t;\mathbf{z},\tau)B_i^\dagger(\mathbf{z},\tau)\right\rangle,
\end{eqnarray}
where $S_F(x,y)$ stands for the charm quark propagator. Accordingly, there are two types of
wall-source quark propagators to be calculated. One of them uses the usual wall source by setting
the source element to unity at each spatial site of the source time slice. The other one uses the
source by multiplying the chromomagnetic field tensor $B_i(\mathbf{z},\tau)$ to each site of the
plain wall source. In order to increase the statistics additionally, for each configuration we
calculate $T$ charm quark propagators $S_F(\vec{x},t;\vec{0},\tau)$ by setting the corresponding
source vectors on each time slice $\tau$. This permits us to average over the temporal direction
when calculating the two-point functions.
\par
In practice, the two-point functions $C^{(H)}(\mathbf{r},t;\tau)$ with the same $r=|\mathbf{r}|$
are averaged, such that the quantum number is kept to be right $J^{PC}=1^{--}$. After averaging
over the time direction, the practical two-point functions we calculate are
\begin{eqnarray}\label{fit_form}
C^{(H)}(r,t)&=&\frac{1}{TN_r}\sum\limits_{|\mathbf{r}|=r}\sum\limits_{\tau=1}^{T}C^{(H)}(\mathbf{r},t+\tau;\tau)\nonumber\\
&=&\sum\limits_i \Phi_i(r)e^{-m_i t},
\end{eqnarray}
where $N_r$ is the degenerate degree of $r=|\mathbf{r}|$. In the data analysis stage, we perform a
simulitaneous multi-exponential fit to $C^{(H)}(r,t)$'s by using a correlated minimal-$\chi^2$ fit
method with the jackknife covariance matrix (we use three mass terms throughout this work).
\begin{figure}
\includegraphics[height=5.0cm]{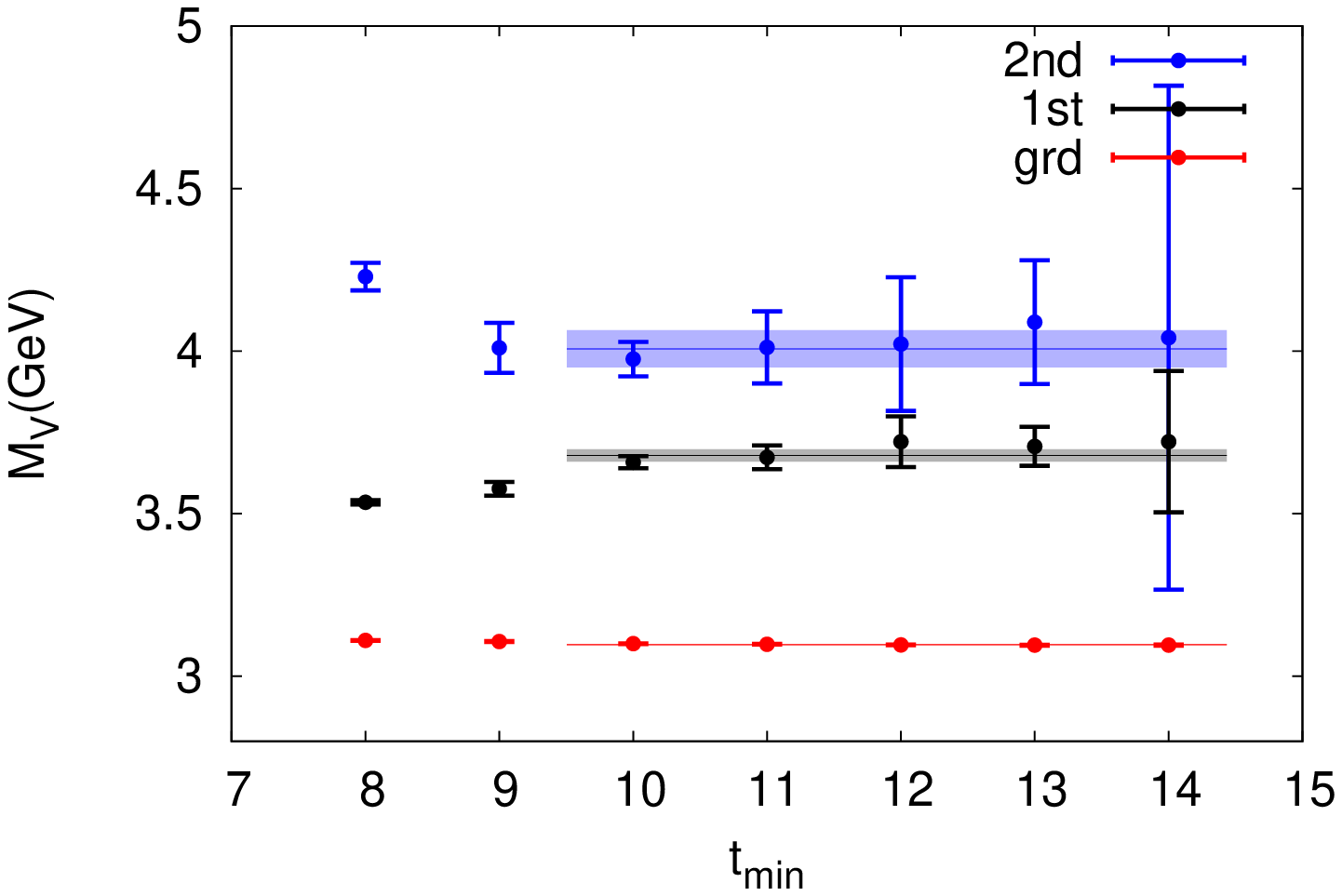}
\includegraphics[height=5.0cm]{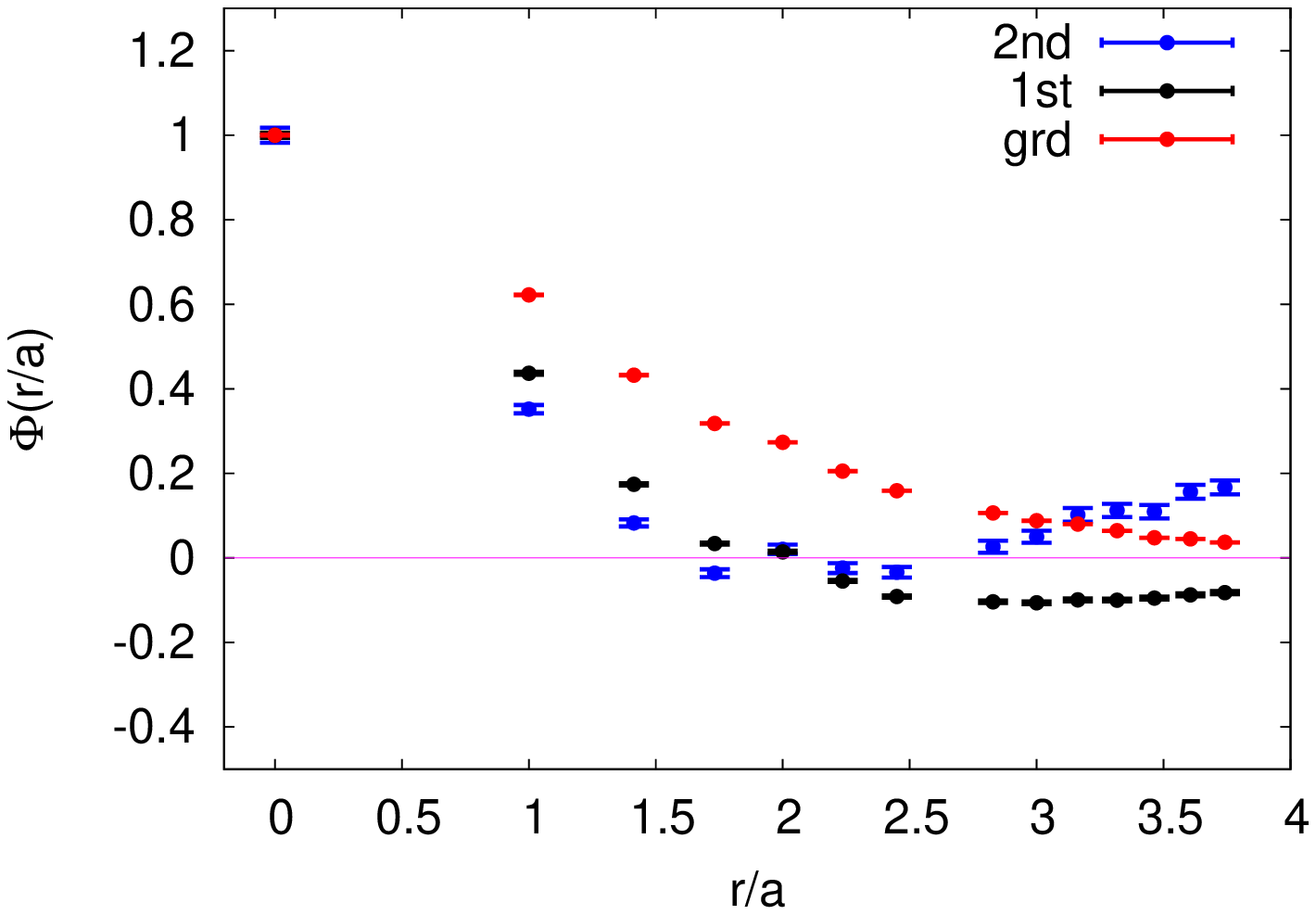}
\caption{ \label{cc}Upper panel: Masses of the three lowest states fitted to $C^{(M)}(r,t)$ with
different $t_{\rm min}$ ($\beta=2.4$). We average the masses in this range with each value weighted
by its error and get the values $m_1=3.097(1)$ GeV, $m_2=3.679(19)$ GeV, and $m_3=4.007(57)$ GeV,
respectively. Lower panel: $\Phi_i'(r/a)$'s (normalized as $\Phi_i'(0)=1$) of the lowest three
states. It is clearly seen that the nodal behaviors are very different for different states. }
\end{figure}

\subsection{Masses and Bethe-Salpeter amplitudes of conventional vector charmonia}
In order to test the reliability of the fitting strategy mentioned above, we first carry out a
similar analysis to the correlation functions $C^{(M)}(r,t)$ of the spatially extended version of
operator $O^{(M)}$ on the $\beta=2.4$ lattice. The procedure is detailed as follows. The spatially
extended version of $O^{(M)}$ is defined as
\begin{equation}
O_i^{(M)}(\mathbf{x},t;\mathbf{r})=\bar{c}(\mathbf{x},\tau)\gamma_i c(\mathbf{x+r},\tau),
\end{equation}
whose correlation function with the corresponding wall source operator
\begin{equation}
O_i^{'(W)}(\tau)=\sum\limits_{\mathbf{y,z}}\bar{c}(\mathbf{y},\tau)\gamma_i c(\mathbf{z},\tau)
\end{equation}
, say, $C^{'(M)}(\mathbf{r},t+\tau;\tau)$, is defined similarly as in Eq.~(\ref{two-point}). After
averaging over the temporal direction, we have
\begin{eqnarray}
C^{'(M)}(r,t)&=&\frac{1}{TN_r}\sum\limits_{|\mathbf{r}|=r}\sum\limits_{\tau=1}^{T}C^{'(M)}
(\mathbf{r},t+\tau;\tau)\nonumber\\
&=&\sum\limits_i \Phi'_i(r)e^{-m_i t}
\end{eqnarray}
where $\Phi'_i(r)$ is the $r$-dependent spectral weight of the $i$-th state.

In the fitting procedure, we fix a maximal $t$ (denoted by $t_{\rm max}$ and varying the lower
bound $t_{\rm min}$ of the fit window, then we obtain the masses of the lowest three states keep
constant to some extent for a series of $t_{\rm min}$, as shown in the upper panel of
Figure~\ref{cc}. We average the masses in this range with each value weighted by its error and get
the values $m_1=3.097(1)$ GeV, $m_2=3.679(19)$ GeV, and $m_3=4.007(57)$ GeV, respectively. These
three states may correspond to $J/\psi$, $\psi'(3686)$, and $\psi(4040)$ and we can almost
reproduce their experimental spectrum. We also plot the $\Phi_i'(r/a_s)$'s (normalized as
$\Phi'_i(0)=1$) in the right panel of Figure~\ref{cc} where one can find that, there is no radial
node for $\Phi'_1(r/a)$, one radial node for $\Phi'_2(r/a_s)$, and there are two radial nodes for
$\Phi'_3(r/a_s)$. Given the quark model assignments $n^3S_1$ state for $J/\psi$, $\psi(3686)$, and
$\psi(4040)$ with $n=1,2,3$, respectively, this is actually not surprising, since $\Phi'_i(r)$ is
proportional to the Coulomb Bethe-Salpeter amplitude of the $i$-th $S$-wave charmonium, which, at
the leading order of the non-relativistic approximaiton, corresponds to the radial wave functions
in the quark model~\cite{Bodwin:1994jh, Chen:2007vu}.

\subsection{Existence of an exotic vector charmonia and its mass}
From above one can see that our data analysis strategy is robust for the conventional vector
charmonium states, therefore we perform the similar study for the correlation functions described
in Eq.~(\ref{two-point}) and (\ref{fit_form}).

\begin{figure}
\includegraphics[height=5.0cm]{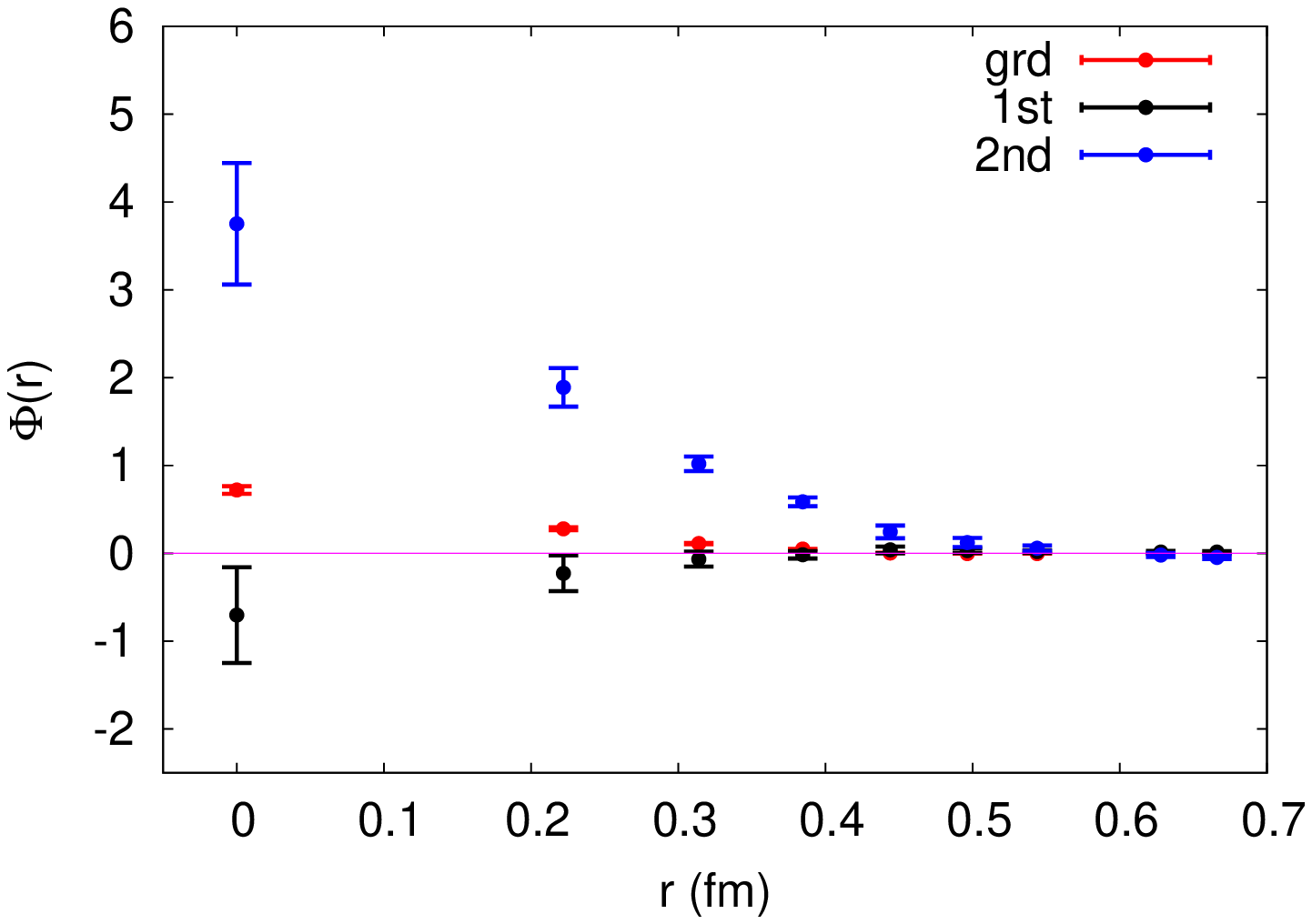}\\
\includegraphics[height=5.0cm]{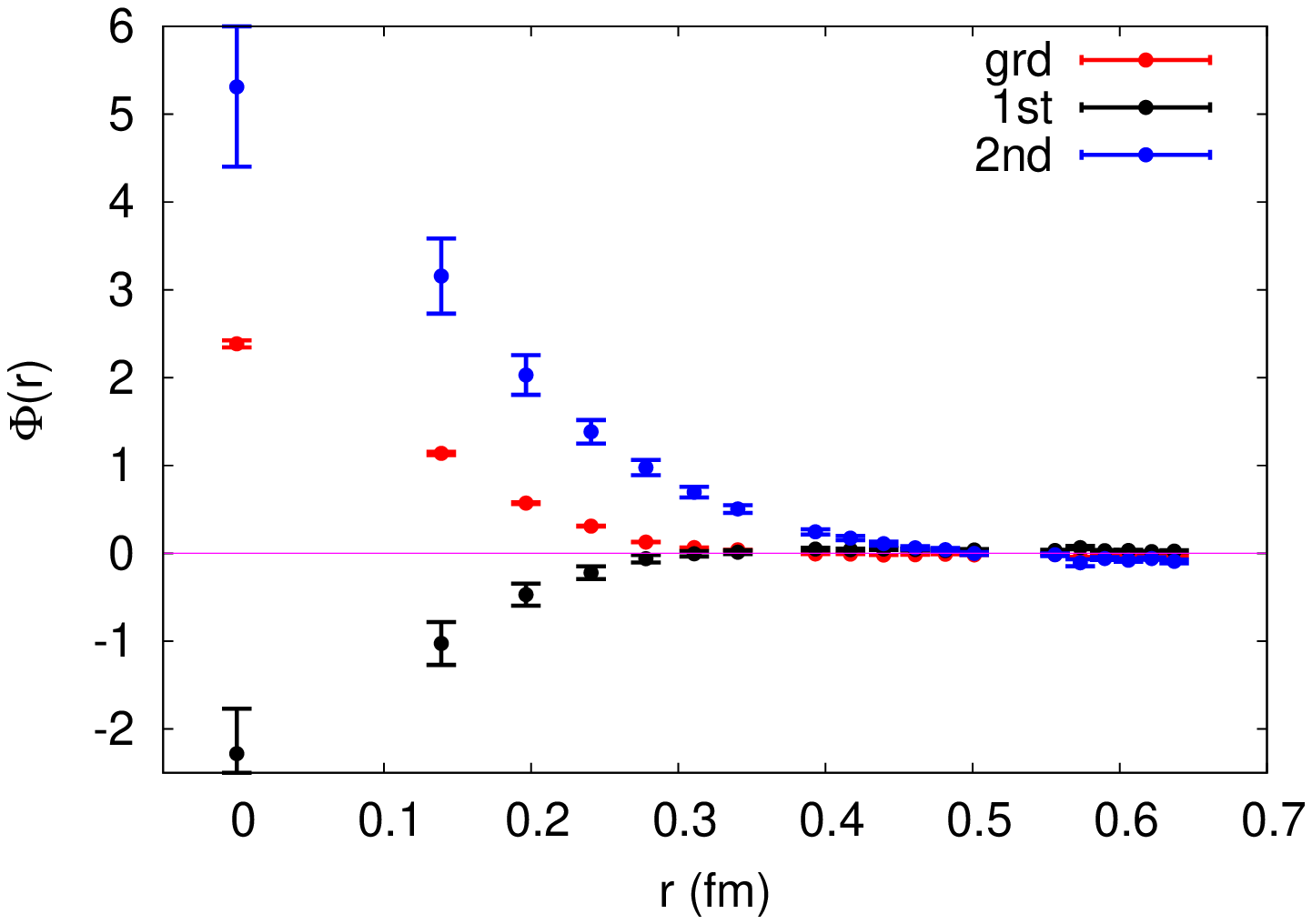}
\caption{ \label{wavefun1}Plots of $\Phi_i(r)$ with respect to $r$ (in phyical units) for the three
lowest states ( the upper panel is for the $\beta=2.4$ case, and the lower panel for $\beta=2.8$
).}
\end{figure}

Figure~\ref{wavefun1} shows the plots of $\Phi_i(r)$ with respect to $r$ (in physical units)
through a three-mass-term fit (the upper panel is for $\beta=2.4$ at $t_{\rm min}=12a_t$, and the
lower panel for $\beta=2.8$), whose masses are fitted to be 3.100(7) GeV, 3.58(9) GeV, and 4.6(2)
GeV for $\beta=2.4$, and 3.090(6) GeV, 3.54(5) GeV, and 4.6(1) GeV for $\beta=2.8$ at $t_{\rm
min}=16a_t$. $\Phi_1(r)$ and $\Phi_2(r)$ damp more rapidly and are close to zero near $r\sim 0.3$
fm while $\Phi_3(r)$ is still relatively large. The lowest two states correspond very possibly to
the conventional vector charmonia $J/\psi$ and $\psi'$ according to their masses. In contrast, the
third state, with a much higher mass, still dominates the two point functions with $r$ larger than
0.3 fm. This may signal the exotic nature of this state that is reflected by the spatially extended
sink operator $O_i^{(H)}$. Of course, the higher conventional vector charmonia, such as
$\psi(4040)$ and $\psi(4415)$, should also contribute to the two-point function $C^{(H)}(r,t)$,
however in our data analysis procedure, $C^{(H)}(r,t)$ cannot accommodate more statistically
meaningful states. The presence of the higher conventional charmonia may result in the small shift
of the masses of the fitted states, for example, the mass of the second state deviates from that of
the would-be $\psi'$ state.

\begin{figure}
\includegraphics[height=5.0cm]{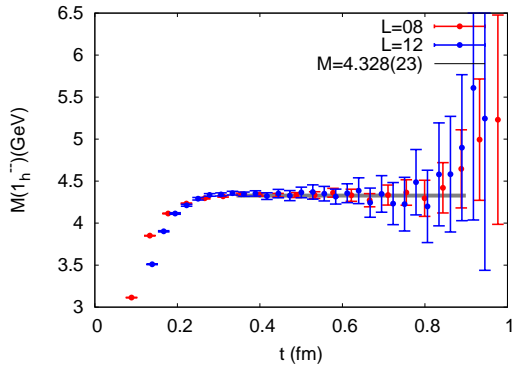}
\caption{ \label{plat}The effective mass plateaus of $C(\omega,t)$'s for $\beta=2.4$ (red points)
and $\beta=2.8$ (blue points). The horizontal line shows the fitted mass $M_X=4.33(2)$ GeV through
a one-exponential fit.}
\end{figure}

It is seen from Fig.~\ref{wavefun1} the $r$-behaviors of the first ($J/\psi$) and the second
state($\psi'$) are similar up to an overall factor. If this is the case for all the conventional
charmonium states since the $r$ behavior may depict the center-of-mass motion of the $c\bar{c}$
component and the conventional charmonia are free of this in the nonrelativistic approximation, we
can conjecture that the spectral weights of the $i$-th conventional charmonia $\Phi_i(r)$ can be
factorized into $\Phi(r)W_i$ where $\Phi(r)$ is approximately uniform and insensitive to the
different conventional charmonia, such that the two-point functions $C^{(H)}(r,t)$ with different
$r$ can be linearly combined to eliminate the contribution from the conventional charmonium states.
In practice, we combine linearly the correlation functions $C^{(H)}(r,t)$ at two specific $r_1$ and
$r_2$ as
\begin{equation}
C(\omega,t)= C^{(H)}(r_1,t)-\omega C^{(H)}(r_2,t)
\end{equation}
where $\omega$ is a tunable parameter. For each lattice, an optimal $\omega$ can be obtained by the
requirement that the effective mass plateau of $C(\omega,t)$ is as long as possible. To be
specific, for $\beta=2.4$, we use $r_1=0$ and $r_2=a_s$ and set $\omega=2.576$, which is very close
to the central value of the ratio $\Phi_1(0)/\Phi_1(a_s)=2.583$ through the three-mass fit
illustrated in Fig~\ref{wavefun1}. Using this omega and the fit results in Fig~\ref{wavefun1}, the
spectral weights of the three states in $C(\omega,t)$ are roughly 0.002, -0.1, and -1.1,
respectively. This implies that the relative contribution of the third state is strongly enhanced
by this subtraction scheme. Similarly, for $\beta=2.8$ we use $r_1=a_s$ and $r_2=\sqrt{3}a_s$ and
set $\omega=3.658$ which is also close to the central value of the ratio
$\Phi_1(a_s)/\Phi_1(\sqrt{3}a_s)=3.681$. The spectral weights of the three states are roughly
0.007, -0.2 and -1.9, respectively.  In Fig.~\ref{plat} we plot the effective mass plateaus of
$C(\omega,t)$'s for $\beta=2.4$ and $\beta=2.8$. The $t$ and the masses are expressed in the
physical units according to the lattice spacings listed in Table~\ref{tab:lattice}. One can see
that both plateaus are fairly good and lie on each other. Since both the sink operator $O^{(H)}(r)$
and the source operator $O^{(W)}$ (whose correlation functions are $C^{(H)}(r,t)$) are expected to
couple strongly to hybrid-like states and the contribution from the conventional charmonia is
subtracted largely by the above scheme, we take the state reflected by the observed plateau as the
exotic vector charmonium and name it as $X$ in the rest part of this work (we keep the name of the
experimental state $X(4260)$). The horizontal line shows the fitted mass $M_X=4.33(2)$ GeV through
a one-exponential fit in the time range from $0.3$ to $0.9$ fm.

To this end, we claim that a vector charmonium-like state with a mass of 4.33(2) GeV has been
unambiguously singled out, whose exotic nature may be reflected by its distinct coupling to the
special interpolation field $O^{(H)}(r,t)$ in comparison with the conventional vector charmonia. It
should be noted that this state has also been observed by previous lattice studies using
variational methods based on lattice operator sets~\cite{Liu:2012ze}, however, the spatially
extended operators $O^{(H)}(r,t)$ we use give more clear picture of its inner structure.

\subsection{Leptonic decay constant of the exotic vector charmonium}

Since this hybrid-like charmonium can be disentangle from the conventional charmonia with the
prescription above, its leptonic decay constant can be investigated accordingly. The leptonic decay
constant $f_V$ of a vector meson state $V$ is defined by
\begin{equation}
\label{psi_decay_constant}
 \langle 0|J_\mu^{\rm (em)}(0)|V(\vec{p},r)\rangle = m_V f_V \epsilon_\mu(\vec{p},r),
\end{equation}
where $J_\mu^{\rm (em)}(0)$ is the electromagnetic current and $\epsilon_\mu(\vec{p},r)$ is the
polarization vector of $V$ at momentum ${\vec p}$. For vector charmonium states, $J_\mu^{\rm
(em)}(0)$ can be approximated by $\bar{c}\gamma_\mu c (0)$ if the contribution from other quark
flavors through annihilation diagrams is neglected. Since the vector current $J_\mu^{\rm em}(x)$
defined in the continuum limit is no longer conserved on the lattice, we perform a nonperturbative
renormalization procedure~\cite{Dudek:2006ej} to extract the multiplicative renormalization
constant $Z_V$ of the current. The renormalization constant of the spatial componets of $J_\mu^{\rm
em}$ is determined to be $Z_V^{(s)}=1.39(2)$ for $\beta=2.4$ and $Z_V^{(s)}=1.11(1)$ for
$\beta=2.8$~\cite{Gui:2012gx}.

Since the spatial components of $J_\mu^{\rm (em)}(x)$ is exactly the normal quark bilinear operator
$O_i^{(M)}$ for vector mesons, the matrix elements in Eq.~(\ref{psi_decay_constant}) can be derived
from the corresponding correlation functions involving the operator $O_i^{(M)}$ along with the
vector current renormalization constant $Z_V^{(s)}$. In order to obtain this matrix elements, we
also calculate other two categories of correlation functions in addition to
$C^{(H)}(\mathbf{r},t)$,
\begin{eqnarray}
C^{(J)}(t)&=&\frac{1}{3}\sum\limits_{\mathbf{x},i}\langle
0|J_i(\mathbf{x},t)O_i^{(W)~\dagger}(0)|0\rangle\nonumber\\
C^{(W)}(t)&=&\frac{1}{3}\sum\limits_{i}\langle 0|O_i^{(W)}(t)O_i^{(W)~\dagger}(0)|0\rangle.
\end{eqnarray}
where averaging over the temporal direction is also taken implicitly in the above expressions.
After the intermediate states insertion to $C^{(J)}(t)$, $C^{H)}(\mathbf{r},t)$, and $C^{(W)}(t)$,
we have
\begin{eqnarray}\label{functions}
C^{(J)}(t)&=&\sum\limits_{n,r}\frac{1}{2m_n}Z_n^{(J)}Z_n^{(W)*}e^{-m_n t},\nonumber\\
C^{(H)}(\mathbf{r},t)&=&\sum\limits_{n,\mathbf{r}}\frac{1}{2m_n}Z_n^{(H)}(r)Z_n^{(W)*}e^{-m_n t},\nonumber\\
C^{(W)}(t)&=&\sum\limits_{n,r}\frac{1}{2m_nV_3}Z_n^{(W)}Z_n^{(W)*}e^{-m_n t},
\end{eqnarray}
where $m_n$ is the mass of the $n$-th state and the parameter $Z_n^{(K)}$ with $K$ refering to $H$
or $W$ is defined as
\begin{equation}
\langle 0|O_i^{(K)}|V_n(\mathbf{p},r)\rangle = Z_n^{(K)}\epsilon_i(\mathbf{p},r).
\end{equation}
Accordingly the leptonic decay constant $f_{V_n}$ can be derived from $Z_n^{(J)}$ from the
definition Eq.~(\ref{psi_decay_constant}) as
\begin{equation}\label{decay_value}
f_{V_n}=C Z_V^{(s)}Z_n^{(J)}/m_n,
\end{equation}
where $C$ is a overall constant prefactor owing to the redefinition of our quark fields and the
anisotropic lattices we are using.

\begin{figure*}
\includegraphics[height=12.0cm]{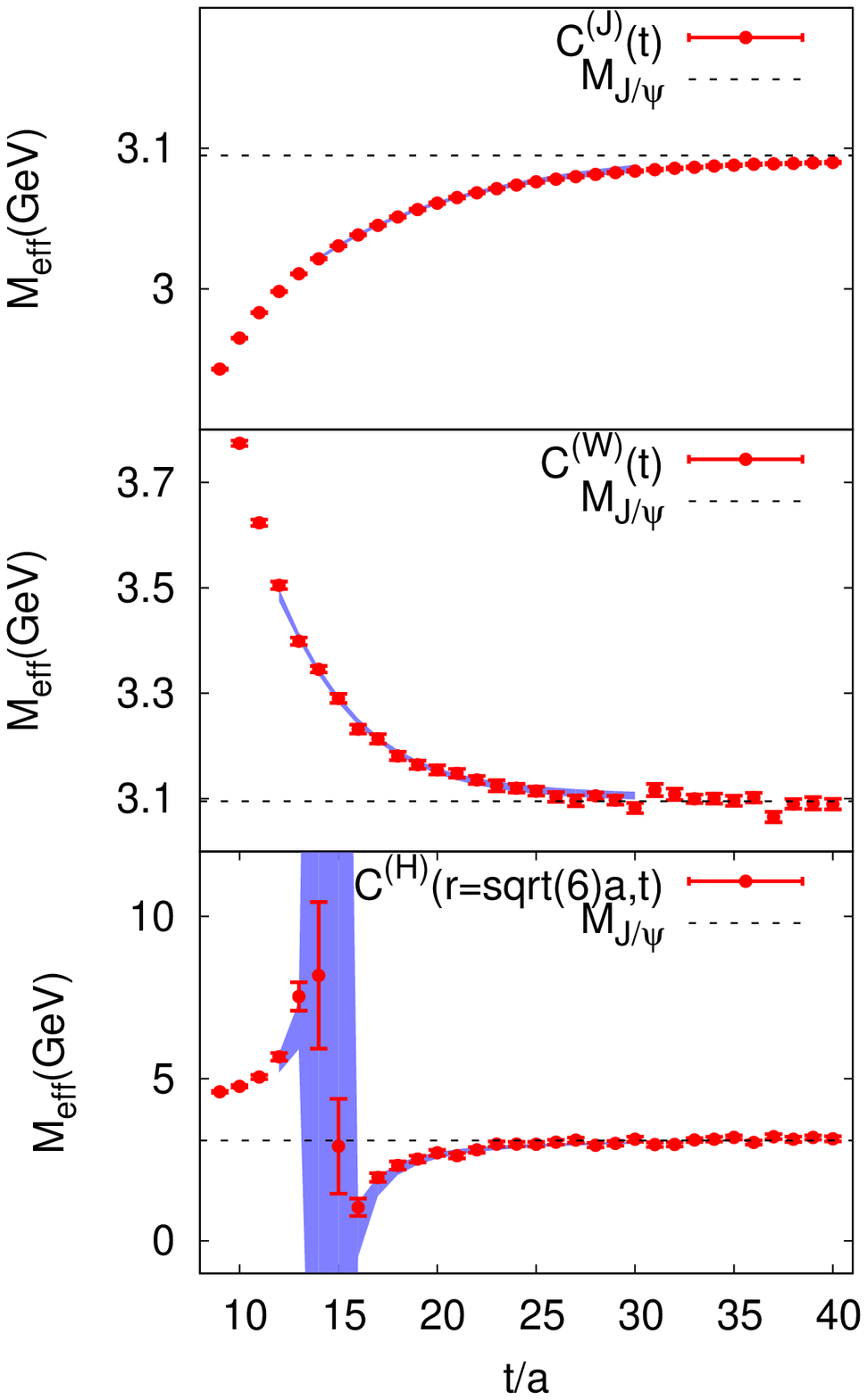}
\includegraphics[height=12.0cm]{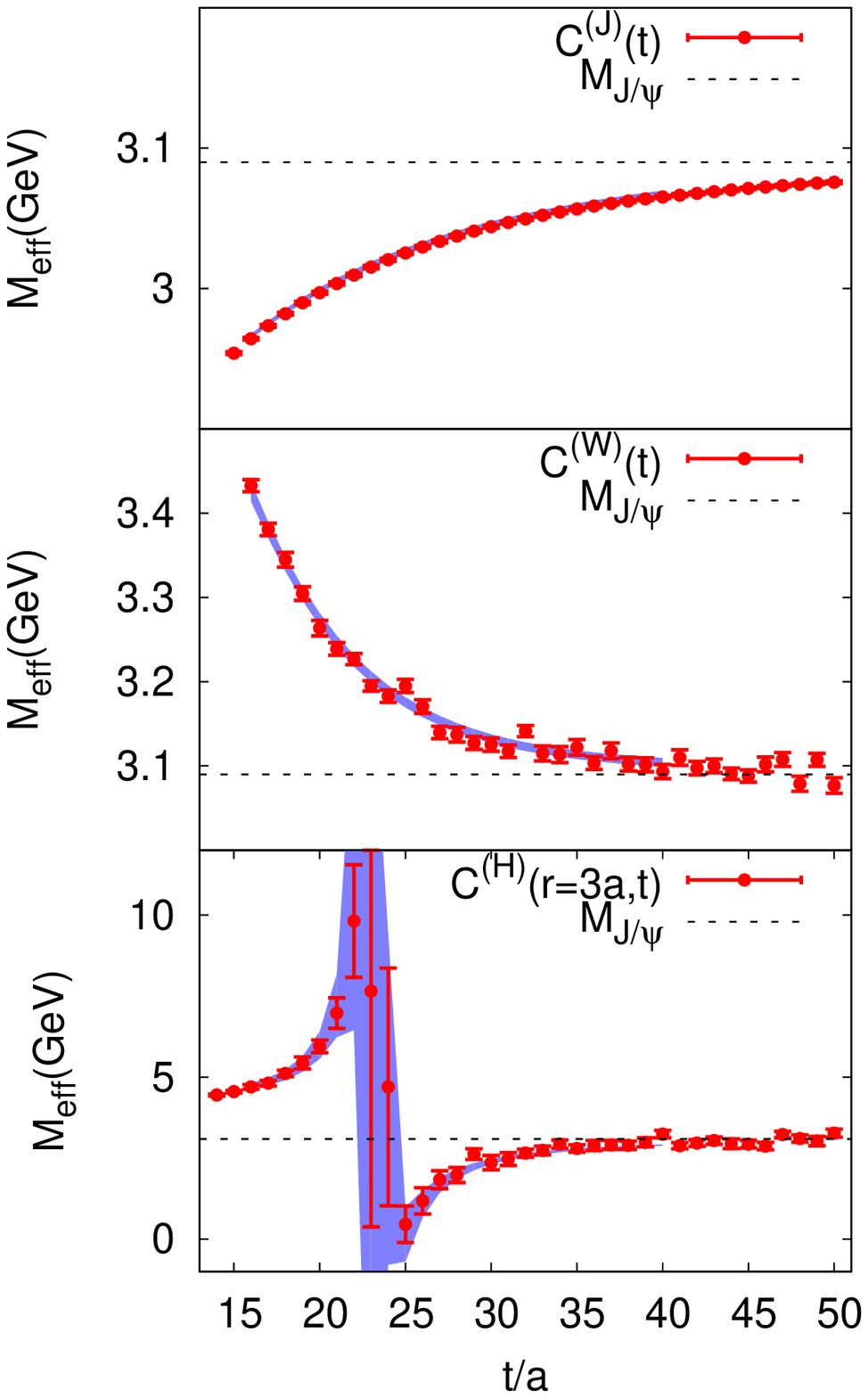}
\caption{ \label{fit_plat} The effective masses $M_{\rm eff}(t)$ (in the physical unit GeV) of the
correlation functions $C^{(X)}(t)$ with $X$ standing for $J$, $H$, and $W$, respectively. The mass
of the ground state ($J/\psi$) is also plotted as a horizontal line to guide eyes. We also show the
effective masses (in blue bands) of the functions in Eq.~(\ref{functions}) using the fitted
parameters at $t_{\rm min}=12a_t$ for $\beta=2.4$ and $t_{\rm min}=16a_t$ for $\beta=2.8$.}
\end{figure*}

The time dependence of the correlation functions is usually observed from their effective mass
plots. The effective masses of different correlation functions $C^{(X)}(t)$ are defined as
\begin{equation}
M_{\rm eff}(t)a=\log\frac{C^{(K)}(t)}{C^{(K)}(t+1)},
\end{equation}
where $K$ stands for $J$, $H$, and $W$ and illustrated in Fig.~\ref{fit_plat}, where the mass of
the ground state ($J/\psi$) is also plotted as a horizontal dashed line to guide eyes. The left
panel of Fig.~\ref{fit_plat} is for the $\beta=2.4$ lattice, and the right panel is for
$\beta=2.8$.   In the plots, we present in the third row the effective mass of $C^{(H)}(r,t)$ at a
specific $r$ ($r=\sqrt{6}a_s$ for $\beta=2.4$ and $r=3a_s$, respectively), where $C^{(H)}(r,t)$ is
dominated by the third state in the short time range, changes the sign and finally saturated by the
ground state when $t$ increases. This is manifested in the effective mass plot by the phenomenon
that $M_{\rm eff}(t)$ shows a meta-stable plateau (roughly $4.4$ GeV) higher than the ground state
before the discontinuity time, and then converges to the mass of the ground state. This phenomenon
also implies that the spectral weight of the third state is much larger than that of the lower
states and the signs of the two spectral weights are different. In the fitting procedure, we fix
the maximal time $t_{\rm max}$ of the fitting window ($t_{\rm max}=30a_t$ for $\beta=2.4$ and
$t_{\rm max}=40a_t$ for $\beta=2.8$), and let the minimal time vary in a range. The fitted spectral
weights of $C^{(J)}(t)$ and $C^{(W)}(t)$ are listed in Table~\ref{fit_results} (raw data), from
which the decay constants can be derived. The fitted masses and the decay constants are converted
to the values in the physical unit (GeV) and are presented in Table~\ref{fit_final}. The spectral
weights of $C^{(H)}(r,t)$ are less relevant and omitted here to save space (one can refer to
Fig.~\ref{wavefun1} to see their relative magnitudes for the three states). All the errors are
statistical and obtained through a jackknife analysis. In order to illustrate the fit quality, we
also show the effective masses (in blue bands) of the functions in Eq.~(\ref{functions}) using the
fitted parameters at $t_{\rm min}=12a_t$ for $\beta=2.4$ and $t_{\rm min}=16a_t$ for $\beta=2.8$.
It is seen that the fit functions describe the measured data very well. Actually for all the fits,
the $\chi^2/N_{\rm d.o.f}$'s are around 1 even though the number of the degree-of-freedom $N_{\rm
d.o.f}$ are always several hundreds.

\begin{table*}[t]
\centering\caption{\label{fit_results}The spectral weights of the three states at different $t_{\rm
min}$ on both lattices through a correlated three-mass-term fit. The errors are obtained through a
jackknife analysis. The experimental results are also listed for comparison.}
\begin{ruledtabular}
\begin{tabular}{cclll|lll}
 $\beta$&$t_{\rm min}$&$|Z_1^{(J)}Z_1^{(W)}|$&$|Z_2^{(J)}Z_2^{(W)}|$&$|Z_3^{(J)}Z_3^{(W)}|$&$|Z_1^{(W)}|^2$& $|Z_2^{(W)}|^2$&$|Z_3^{(W)}|^2$\\
        &             &                      &                      &                      &
        $(\times10^5)$&       $(\times10^5)$            &$(\times10^6)$ \\
 \hline
        &   11        &  14.3(0.6) &    9.7(0.3) &  1.3(1.9)  &   0.144(6)  &   0.08(5)  &  0.41(8) \\
        &   12        &  14.4(0.8) &    9.5(0.5) &  2.0(2.7)  &   0.146(7)  &   0.06(7)  &  0.33(9) \\
   2.4  &   13        &  13.9(0.4) &    9.9(0.7) &  0.1(3.2)  &   0.142(3)  &   0.12(5)  &  0.39(17)\\
        &   14        &  13.8(0.4) &   10.6(1.8) &  3.3(7.2)  &   0.141(3)  &   0.15(9)  &  0.39(24)\\
        &   15        &  13.6(0.4) &   11.6(2.5) &  8.0(8.7)  &   0.140(2)  &   0.18(8)  &  0.33(24)\\
 \hline
        &   14        &  33.6(2.8) &   22.6(1.9) &  0.3(2.1)  &   3.86(27)  &  0.88(0.97)& 5.9(1.0)\\
        &   15        &  31.8(2.2) &   21.6(1.1) &  1.7(3.0)  &   3.65(22)  &  1.39(1.28)& 4.3(1.1)\\
  2.8   &   16        &  32.1(1.6) &   21.5(0.9) &  1.3(2.3)  &   3.68(14)  &  1.39(0.69)& 3.9(0.7)\\
        &   17        &  33.1(4.0) &   22.0(1.6) &  0.0(5.2)  &   3.81(38)  &  0.66(1.95)& 3.4(0.9)\\
        &   18        &  31.5(2.0) &   20.7(1.1) &  0.4(6.1)  &   3.64(15)  &  1.95(0.91)& 7.7(2.9)\\

\end{tabular}
\end{ruledtabular}
\end{table*}

\begin{table*}[t]
\caption{\label{fit_final} The masses and the leptonic decay constants of the three states at
different $t_{\rm min}$ on both lattices. All the values are in GeV using the lattice spacings
listed in Table~\ref{tab:lattice}. The errors are obtained through a jackknife analysis. The
experimental results are also listed for comparison.}
\begin{ruledtabular}
\begin{tabular}{ccllllll}
 $\beta$&$t_{\rm min}$&$m_{J/\psi}$&$f_{J/\psi}$& $m_{\psi'}$&$f_{\psi'}$& $m_3$  & $f_3$ \\
 \hline
        &   11        &  3.101(6)  &  0.47(2)   &  3.60(7)  &  0.4(1)   &  4.6(1)&-0.005(8)\\
        &   12        &  3.100(7)  &  0.47(2)   &  3.58(9)  &  0.4(2)   &  4.6(2)&-0.009(10)\\
   2.4  &   13        &  3.097(4)  &  0.46(1)   &  3.64(6)  &  0.31(6)  &  4.7(2)& 0.00(1)\\
        &   14        &  3.096(4)  &  0.46(1)   &  3.68(9)  &  0.29(6)  &  4.7(3)& 0.01(3)\\
        &   15        &  3.095(4)  &  0.46(1)   &  3.72(9)  &  0.29(5)  &  4.7(3)& 0.04(4)\\
 \hline
        &   14        &  3.096(10) &  0.44(2)   &  3.51(7)  &  0.5(4)   &  4.7(1)& 0.001(5)\\
        &   15        &  3.090(9)  &  0.43(2)   &  3.56(8)  &  0.4(2)   &  4.6(2)& 0.005(7)\\
  2.8   &   16        &  3.090(6)  &  0.43(1)   &  3.54(5)  &  0.4(1)   &  4.6(1)& 0.004(6)\\
        &   17        &  3.090(10) &  0.43(3)   &  3.50(10) &  0.6(8)   &  4.5(2)& 0.00(2)\\
        &   18        &  3.088(8)  &  0.42(2)   &  3.55(7)  &  0.3(9)   &  4.9(2)& 0.01(1)\\
 \hline\\
        &  Expt.      &  3.097     &  0.407(5)  &  3.686    &  0.290(2) &        &        \\
\end{tabular}
\end{ruledtabular}
\end{table*}

As shown in Table~\ref{fit_results} and Table~\ref{fit_final}, the fitted parameters are almost
stable and insensitive to $t_{\rm min}$. The masses of the first and the second states are
consistent with the those of $J/\psi$ and $\psi'$, while the mass of the third state is a little
higher than the hybrid-like state we obtain before. This can be attributed to the contamination
from even higher states to some extent. In the right part of Table~\ref{fit_results}, the spectral
weights $|Z_3^{(W)}|^2$ are an order of magnitude larger than those of the lowest two states
(corresponding mostly to $J/\psi$ and $\psi'$). This is not strange since $Z_n^{(W)}$ is the
coupling of the hybrid-like wall-source operator $O^{(W)}$ to the $n$-th state and is expected to
be enhanced when coupling to a hybrid-like state. In contrast, the spectral weights
$|Z_n^{(J)}Z_n^{(W)}|$ (in the left part of Table~\ref{fit_results}) of the lowest two states, even
suppressed by $Z_n^{(W)}$, are much larger than that of the third state which are close to zero
with errors. This may imply that the decay constant of the third state is very small, since
$Z_n^{(J)}$ is proportional to the decay constant of the $n$-th state. Using the fitted spectral
weights $Z_n^{(J)}Z_n^{(W)}$ and $|Z_n^{(W)}|^2$, we can get the concrete values of $Z_n^{(J)}$,
from which the decay constant of the $n$-th state can be derived using Eq.~(\ref{decay_value}). The
derived decay constants of the three states are also listed in Table~\ref{fit_final}, where the
last row lists the experimental values for comparison. For $J/\psi$, we get its decay constant to
be roughly 0.46(2) GeV at $\beta=2.4$ and 0.43(2)GeV at $\beta=2.8$, which are close to the
experimental value although 5\%-10\% larger. The deviation can be attributed to the artifact of the
finite lattice spacings (and also the uncertainty owing to the quenched approximation). The derived
decay constant of $\psi'$, $f_{\psi'}$, seems compatible with the experimental value, but with huge
errors which come mainly from the uncertainty of $|Z_2^{(W)}|^2$.

The most striking observation is that the decay constant $f_3$ of the third state is consistent
with zero within the error. Superficially, it seems that the exotic vector charmoium has a nearly
zero decay constant. However, there is a possibility that this is a mixing effect of two nearby
states (for example, the would-be exotic state and $\psi(4415)$), whose contribution to
$C^{(J)}(t)$ cancels to some extent, because we perform the simultaneous fit using only three mass
terms. We have addressed that an exotic vector charmoninum state does exist and contributes
substantially to the correlation functions $C^{(H)}(r,t)$ in previous context. This is also the
case for $C^{(W)}(t)$ since $|Z_3^{(W)}|^2$ is one magnitude or even more larger than
$|Z_1^{(W)}|^2$ and $|Z_2^{(W)}|^2$. Based on these facts and considering the possible admixture of
a conventional charmonium state to the exotic state, we try to estimate the upper limit of the
decay constant of the exotic vector chrmonium. If the third state is actually contributed from the
would-be exotic state $X$ and the adjacent vector charmonium state $\psi(4415)$, then the spectral
weight $Z^{(J)}_3 Z^{(W)}_3$ can be expressed as
\begin{equation}
Z^{(J)}_3 Z^{(W)}_3= Z^{(J)}_X Z^{(W)}_X + Z^{(J)}_{\psi(4415)} Z^{(W)}_{\psi(4415)}\sim 0.
\end{equation}
Thus we have
\begin{equation}
|Z^{(J)}_X|\sim \frac{|Z^{(W)}_{\psi{4415}}|}{|Z^{(W)}_X|}|Z^{(J)}_{\psi(4415)}|.
\end{equation}
According to Eq.~(\ref{decay_value}), this is equivalent to
\begin{equation}
f_X\sim \frac{|Z^{(W)}_{\psi(4415)}|}{|Z^{(W)}_X|} f_{\psi(4415)}.
\end{equation}
Since $|Z_3^{(W)}|^2\gg |Z_1^{(W)}|^2\sim |Z_2^{(W)}|^2$, we can take the approximation
$Z_3^{(W)}\approx Z_X^{(W)}$. Furthermore, if we assume $|Z^{(W)}_{\psi(4415)}|\sim |Z^{(W)}_2|$,
from Table~\ref{fit_results}, we can take $Z^{(W)}_{\psi(4415)}/Z^{(W)}_X\sim 1/5$. Experimentally,
the leptonic decay width of $\psi(4415)$ is measured to be $\Gamma(\psi(4415)\rightarrow
e^+e^-=0.58(7)$keV, $f_{\psi(4415)}$ is extracted to be 157 MeV using the relation.
\begin{equation}
\Gamma(V_{c\bar{c}}\rightarrow e^+e^-)=\frac{16\pi}{27}\alpha_{\rm QED}^2 \frac{f_V^2}{M_V}.
\end{equation}
where we take $\alpha_{\rm QED}=1/134$ at the charm quark mass scale. Therefore, $f_X$ can be
roughly estimated to be
\begin{equation}
f_X\sim 30\,{\rm MeV}.
\end{equation}
So we can assign a safer upper limit of $f_X$ as
\begin{equation}
f_X<\frac{1}{10}f_{J/\psi}\sim 40~~\,{\rm MeV},
\end{equation}
which gives the upper limit of the leptonic decay width of the exotic vector charmonium,
\begin{equation}
\Gamma(X\rightarrow e^+e^-)< 40\,{\rm eV}.
\end{equation}

\section{Discussion}

The suprisingly small $e^+e^-$ decay width of the exotic vector charmonium $X$ is in sharply
contrast to that of conventional vector charmonia, which are usually of the order of keV. In other
words, if this hybrid-like vector charmonium does exist in the real world, its contribution to the
inclusive cross section of $e^+e^-$ annihilation is rather small. Actually the $R$ value scan
versus the invariant energy $\sqrt{s}$ of $e^+e^-$ collision does not show any indication of an
extra vector charmonium-like state around $\sqrt{s}=4.3$ GeV (there is however a small dip in this
energy range). The BEPCII/BESIII in Beijing is now accumulating the data of $e^+e-$ collision in
this energy range and will hopefully give more precise line shape of $R$-value here to clarify the
situation. On the other hand, as mentioned above, the vector charmonium state $X(4260)$ was
observed by several experiments in the initial state radiation of $e^+e^-$ annihilation into
$J/\psi \pi^+\pi^-$. The combined decay width of $X(4260)$ is
\begin{equation}\label{combined_width}
\Gamma(X(4260)\rightarrow e^+e^-)Br(X(4260)\rightarrow J/\psi\pi\pi)=9.2\pm 1.0{\rm eV}.
\end{equation}
If $X(4260)$ is tentatively assigned to the $X$ state investigated in this study, combining the
above value with the leptonic decay width of $X$, we can give an estimate of the branch ratio of
$X(4260)$ decaying into $J/\psi \pi^+\pi^-$
\begin{equation}
Br(X(4260)\rightarrow J/\psi\pi\pi)> 20 \%,
\end{equation}
which means that $J/\psi\pi\pi$ is one of the most important decay mode of $X(4260)$. This can
naturally explain why $X(4260)$ was only observed in this channel till now. Furthermore, given the
likely hybrid nature of the $X$ state, it can be expected that the spin singlet $c\bar{c}$
component of $X$ prefers a hadronic transition into spin singlet charmonium, such as $h_c$. So
$X(4260)\rightarrow h_c\pi\pi$ can be also an important decay mode of $X(4260)$. Recently the
BESIII Collaboration studied the $e^+e^-\rightarrow \pi^+\pi^- h_c$ process at the center-of-mass
energies from 3.90 GeV to 4.42 GeV. They found that the cross sections are of the same order of
magnitude as, but have different line shape from those of $e^+e-\rightarrow \pi^+\pi^-
J/\psi$~\cite{Ablikim:2013wzq}. Anyway, it is highly desired to investigate whether they are from
the same resonance structure or not.
\par
The reason for the large branching ratio of $X(4260)\rightarrow J/\psi\pi\pi$ can be depicted as
follows. In the $e^+e^-$ annihilation, the charm quark-antiquark pair $c\bar{c}$ is produced in the
short range through the virtual photon. During the hadronization procedure, the charm quark and the
charm antiquark emit soft gluons continuously, which form a colored gluon halo around the gradually
localized color octet $c\bar{c}$ (in a relative sense). Finally a meta-stable state is formed as
the $X(4260)$ particle. Obviously, the color octet $c\bar{c}$ kernel can be readily neutralized
into a color singlet charmonium by absorbing (emitting) soft gluons from (to) the halo, and the
gluon halo thereby becomes color neutral and are emitted as light hadrons, for example, the
$\pi\pi$ pair. If this is actually the case, the color flux between the charm quark and antiquark
have less chance to be excited to a high enough energy to break, and thus the possibility of the
$D\bar{D}$ decay modes are suppressed. There is a little similarity between this 'halo charmonium'
picture and the so-called 'hadro-charmonium' picture, where the relatively localized color neutral
$c\bar{c}$ kernel is surrounded by a light hadron cloud~\cite{Dubynskiy:2008mq}. However, the
advantage of the former resides in that the direct color interaction between the halo and the
kernel provides an obvious binding mechanism, while in the 'hadron-charmonium picture, there need
more phenomenological assumptions to describe the interaction between the meson cloud and the
charmonium kernel.

\section{Conclusion}

To summarize, we use a new type of spatially extended hybrid-like operator to investigate the
possible existence of exotic vector charmonia. In the non-relativistic approximation of these
operators, the localized color octet charm quark-antiquark component is in the spin singlet state
and separates from the chromo-magnetic field strength with a spatial distance. These operators
couple preferably to a higher vector state $X$ with a mass of 4.33(2) GeV when the distance
increases. This observation indicates that the charm quark-antiquark pair of $X$ may acquire a
center-of-mass motion by recoiling against additional degrees of freedom depicted by the
chromo-magnetic field strength operator, which are necessarily gluonic in the quenched
approximation. In this meaning, the state $X$ can be taken as a hybrid-like vector charmonium. In
addition, through a simultaneous multi-state fit to different correlation functions built from the
vector current operator and the the hybrid operator mentioned above, the leptonic decay constant of
$X$ is tentatively determined to be roughly one order of magnitude smaller than $f_{J/\psi}$, say,
$f_X<40$ MeV, which gives a very small leptonic decay width $\Gamma(X\rightarrow e^+ e^-)<40$ eV.
This is a very important characteristic parameter for $X$ to be identified from experiments.
Obviously the mass and the leptonic decay width of $X$ are consistent with the production and decay
properties of $X(4260)$, which escapes so far from the direct measure in the $e^+e^-$ annihilation.
Based on the combined width $\Gamma_{ee}\Gamma_{J/\psi\pi^+\pi^-}/\Gamma_{\rm tot}=9.2\pm1.0$ eV of
$X(4260)$, if it can be assigned to the $X$ state in this study, its decay branch fraction of
$J/\psi\pi\pi$ mode can be larger than 20\%, which also naturally explains why $X(4260)$ is
dominantly observed in $J/\psi\pi\pi$. By virtue of the inner structure of $X$, $X(4260)$ should
also be observed in the $h_c\pi\pi$ channel.

\section*{Acknowledgments}
The numerical calculations are carried out on Tianhe-1A at the National Supercomputer Center (NSCC)
in Tianjin and the GPU cluster at Hunan Normal University. This work is supported in part by the
National Science Foundation of China (NSFC) under Grants No. 11575196, No. 11575197, No. 11335001,
and 11405053. Y.C. and Z.L. also acknowledge the support of NSFC under No. 11261130311 (CRC 110 by
DFG and NSFC).


\begin{thebibliography}{90}

\vspace{3mm}

\bibitem{Aubert:2005rm}
  B.~Aubert {\it et al.} [BaBar Collaboration],
  %``Observation of a broad structure in the $\pi^+ \pi^- J/\psi$ mass spectrum around 4.26-GeV/c$^2$,''
  Phys.\ Rev.\ Lett., {\bf 95}: 142001 (2005)
%  doi:10.1103/PhysRevLett.95.142001
  [hep-ex/0506081]
  %%CITATION = doi:10.1103/PhysRevLett.95.142001;%%
  %613 citations counted in INSPIRE as of 25 Mar 2016
%\cite{Yuan:2007sj}
\bibitem{Yuan:2007sj}
  C.~Z.~Yuan {\it et al.} [Belle Collaboration],
  %``Measurement of e+ e- ---> pi+ pi- J/psi cross-section via initial state radiation at Belle,''
  Phys.\ Rev.\ Lett.,  {\bf 99}: 182004 (2007)
%  doi:10.1103/PhysRevLett.99.182004
  [arXiv:0707.2541 [hep-ex]]
  %%CITATION = doi:10.1103/PhysRevLett.99.182004;%%
  %303 citations counted in INSPIRE as of 25 Mar 2016
%\cite{Coan:2006rv}
\bibitem{Coan:2006rv}
  T.~E.~Coan {\it et al.} [CLEO Collaboration],
  %``Charmonium decays of Y(4260), psi(4160) and psi(4040),''
  Phys.\ Rev.\ Lett., {\bf 96}: 162003 (2006)
 % doi:10.1103/PhysRevLett.96.162003
  [hep-ex/0602034]
  %%CITATION = doi:10.1103/PhysRevLett.96.162003;%%
  %200 citations counted in INSPIRE as of 25 Mar 2016
%\cite{Agashe:2014kda}
\bibitem{Agashe:2014kda}
  K.~A.~Olive {\it et al.} [Particle Data Group Collaboration],
  %``Review of Particle Physics,''
  Chin.\ Phys.\ C, {\bf 38}: 090001 (2014)
%  doi:10.1088/1674-1137/38/9/090001
  %%CITATION = doi:10.1088/1674-1137/38/9/090001;%%
  %3413 citations counted in INSPIRE as of 25 Mar 2016
%\cite{Zhu:2005hp}
\bibitem{Zhu:2005hp}
  S.~L.~Zhu,
  %``The Possible interpretations of Y(4260),''
  Phys.\ Lett., B {\bf 625}: 212 (2005)
%  doi:10.1016/j.physletb.2005.08.068
  [hep-ph/0507025]
  %%CITATION = doi:10.1016/j.physletb.2005.08.068;%%
  %160 citations counted in INSPIRE as of 12 Apr 2016

%\cite{Close:2005iz}
\bibitem{Close:2005iz}
  F.~E.~Close and P.~R.~Page,
  %``Gluonic charmonium resonances at BaBar and BELLE?,''
  Phys.\ Lett., B {\bf 628}: 215 (2005)
 % doi:10.1016/j.physletb.2005.09.016
  [hep-ph/0507199]
  %%CITATION = doi:10.1016/j.physletb.2005.09.016;%%
  %197 citations counted in INSPIRE as of 12 Apr 2016

%\cite{Kou:2005gt}
\bibitem{Kou:2005gt}
  E.~Kou and O.~Pene,
  %``Suppressed decay into open charm for the Y(4260) being an hybrid,''
  Phys.\ Lett.,  B {\bf 631}: 164 (2005)
  %doi:10.1016/j.physletb.2005.09.013
  [hep-ph/0507119]
  %%CITATION = doi:10.1016/j.physletb.2005.09.013;%%
  %161 citations counted in INSPIRE as of 12 Apr 2016

%\cite{Lacock:1996vy}
\bibitem{Lacock:1996vy}
  P.~Lacock {\it et al.} [UKQCD Collaboration],
  %``Orbitally excited and hybrid mesons from the lattice,''
  Phys.\ Rev.\ D, {\bf 54}: 6997 (1996)
%  doi:10.1103/PhysRevD.54.6997
  [hep-lat/9605025]
  %%CITATION = doi:10.1103/PhysRevD.54.6997;%%
  %118 citations counted in INSPIRE as of 12 Apr 2016

%\cite{Bernard:1997ib}
\bibitem{Bernard:1997ib}
  C.~W.~Bernard {\it et al.} [MILC Collaboration],
  %``Exotic mesons in quenched lattice QCD,''
  Phys.\ Rev.\ D, {\bf 56}: 7039 (1997)
 % doi:10.1103/PhysRevD.56.7039
  [hep-lat/9707008]
  %%CITATION = doi:10.1103/PhysRevD.56.7039;%%
  %173 citations counted in INSPIRE as of 12 Apr 2016

%\cite{Liao:2002rj}
\bibitem{Liao:2002rj}
  X.~Liao and T.~Manke,
  %``Excited charmonium spectrum from anisotropic lattices,''
  hep-lat/0210030
  %%CITATION = HEP-LAT/0210030;%%
  %88 citations counted in INSPIRE as of 12 Apr 2016

%\cite{Bernard:2003jd}
\bibitem{Bernard:2003jd}
  C.~Bernard {\it et al.},
  %``Lattice calculation of 1-+ hybrid mesons with improved Kogut-Susskind fermions,''
  Phys.\ Rev.\ D, {\bf 68}: 074505 (2003)
 % doi:10.1103/PhysRevD.68.074505
  [hep-lat/0301024]
  %%CITATION = doi:10.1103/PhysRevD.68.074505;%%
  %63 citations counted in INSPIRE as of 12 Apr 2016

%\cite{Mei:2002ip}
\bibitem{Mei:2002ip}
  Z.~H.~Mei and X.~Q.~Luo,
  %``Exotic mesons from quantum chromodynamics with improved gluon and quark actions on the anisotropic lattice,''
  Int.\ J.\ Mod.\ Phys.\ A, {\bf 18}: 5713 (2003)
 % doi:10.1142/S0217751X03017038
  [hep-lat/0206012]
  %%CITATION = doi:10.1142/S0217751X03017038;%%
  %59 citations counted in INSPIRE as of 12 Apr 2016

%\cite{Dudek:2009qf}
\bibitem{Dudek:2009qf}
  J.~J.~Dudek, R.~G.~Edwards, M.~J.~Peardon, D.~G.~Richards and C.~E.~Thomas,
  %``Highly excited and exotic meson spectrum from dynamical lattice QCD,''
  Phys.\ Rev.\ Lett., {\bf 103}: 262001 (2009)
%  doi:10.1103/PhysRevLett.103.262001
  [arXiv:0909.0200 [hep-ph]]
  %%CITATION = doi:10.1103/PhysRevLett.103.262001;%%
  %110 citations counted in INSPIRE as of 12 Apr 2016

%\cite{Liu:2012ze}
\bibitem{Liu:2012ze}
  L.~Liu {\it et al.} [Hadron Spectrum Collaboration],
  %``Excited and exotic charmonium spectroscopy from lattice QCD,''
  JHEP, {\bf 1207}: 126 (2012)
%  doi:10.1007/JHEP07(2012)126
  [arXiv:1204.5425 [hep-ph]]
  %%CITATION = doi:10.1007/JHEP07(2012)126;%%
  %124 citations counted in INSPIRE as of 12 Apr 2016

%\cite{Yang:2012gz}
\bibitem{Yang:2012gz}
  Y.~B.~Yang, Y.~Chen, G.~Li and K.~F.~Liu,
  %``Is $1^-+$ Meson a Hybrid?,''
  Phys.\ Rev.\ D, {\bf 86}: 094511 (2012)
 % doi:10.1103/PhysRevD.86.094511
  [arXiv:1202.2205 [hep-ph]]
  %%CITATION = doi:10.1103/PhysRevD.86.094511;%%
  %1 citations counted in INSPIRE as of 05 Apr 2016

%\cite{Dudek:2008sz}
\bibitem{Dudek:2008sz}
  J.~J.~Dudek and E.~Rrapaj,
  %``Charmonium in lattice QCD and the non-relativistic quark-model,''
  Phys.\ Rev.\ D, {\bf 78}: 094504 (2008)
%  doi:10.1103/PhysRevD.78.094504
  [arXiv:0809.2582 [hep-ph]]
  %%CITATION = doi:10.1103/PhysRevD.78.094504;%%
  %34 citations counted in INSPIRE as of 12 Apr 2016

%\cite{Dudek:2009kk}
\bibitem{Dudek:2009kk}
  J.~J.~Dudek, R.~Edwards and C.~E.~Thomas,
  %``Exotic and excited-state radiative transitions in charmonium from lattice QCD,''
  Phys.\ Rev.\ D, {\bf 79}: 094504 (2009)
%  doi:10.1103/PhysRevD.79.094504
  [arXiv:0902.2241 [hep-ph]]
  %%CITATION = doi:10.1103/PhysRevD.79.094504;%%
  %70 citations counted in INSPIRE as of 12 Apr 2016

\bibitem{Foldy:1950} L.L Foldy and S.A. Wouthuysen, Phys. Rev., {\bf 78}: 29 (1950);
                     S. Tani, Prog. Theor. Phys., {\bf 6}: 267 (1951)


\bibitem{Morningstar:1997ff}
  C.~J.~Morningstar and M.~J.~Peardon,
  %``Efficient glueball simulations on anisotropic lattices,''
  Phys.\ Rev.\ D, {\bf 56}: 4043 (1997)
%  doi:10.1103/PhysRevD.56.4043
  [hep-lat/9704011]
  %%CITATION = doi:10.1103/PhysRevD.56.4043;%%
  %296 citations counted in INSPIRE as of 12 Apr 2016
%\cite{Morningstar:1999rf}
\bibitem{Morningstar:1999rf}
  C.~J.~Morningstar and M.~J.~Peardon,
  %``The Glueball spectrum from an anisotropic lattice study,''
  Phys.\ Rev.\ D, {\bf 60}: 034509 (1999)
 % doi:10.1103/PhysRevD.60.034509
  [hep-lat/9901004]
  %%CITATION = doi:10.1103/PhysRevD.60.034509;%%
  %707 citations counted in INSPIRE as of 12 Apr 2016
%\cite{Chen:2005mg}
\bibitem{Chen:2005mg}
  Y.~Chen {\it et al.},
  %``Glueball spectrum and matrix elements on anisotropic lattices,''
  Phys.\ Rev.\ D, {\bf 73}: 014516 (2006) 014516
%  doi:10.1103/PhysRevD.73.014516
  [hep-lat/0510074]
  %%CITATION = doi:10.1103/PhysRevD.73.014516;%%
  %315 citations counted in INSPIRE as of 12 Apr 2016
%\cite{Liu:2001ss}
\bibitem{Liu:2001ss}
  C.~Liu, J.~h.~Zhang, Y.~Chen and J.~P.~Ma,
  %``Calculating the I = 2 pion scattering length using tadpole improved clover Wilson action on coarse anisotropic lattices,''
  Nucl.\ Phys.\ B, {\bf 624}: 360 (2002)
 % doi:10.1016/S0550-3213(01)00662-9
  [hep-lat/0109020]
  %%CITATION = doi:10.1016/S0550-3213(01)00662-9;%%
  %55 citations counted in INSPIRE as of 12 Apr 2016
%\cite{Liu:2005tc}
\bibitem{Liu:2005tc}
  L.~m.~Liu, S.~q.~Su, X.~Li and C.~Liu,
  %``Quenched charmed meson spectra using tadpole improved quark action on anisotropic lattices,''
  Chin.\ Phys.\ Lett., {\bf 22}: 2198 (2005)
 % doi:10.1088/0256-307X/22/9/016
  [hep-lat/0505006]
  %%CITATION = doi:10.1088/0256-307X/22/9/016;%%
  %8 citations counted in INSPIRE as of 12 Apr 2016
%\cite{Yang:2012mya}
\bibitem{Yang:2012mya}
  Y.~B.~Yang {\it et al.} [CLQCD Collaboration],
  %``Lattice study on $\eta_{c2}$ and X(3872),''
  Phys.\ Rev.\ D, {\bf 87}: 014501 (2013)
%  doi:10.1103/PhysRevD.87.014501
  [arXiv:1206.2086 [hep-lat]]
  %%CITATION = doi:10.1103/PhysRevD.87.014501;%%
  %17 citations counted in INSPIRE as of 12 Apr 2016
%\cite{Bodwin:1994jh}
\bibitem{Bodwin:1994jh}
  G.~T.~Bodwin, E.~Braaten and G.~P.~Lepage,
  %``Rigorous QCD analysis of inclusive annihilation and production of heavy quarkonium,''
  Phys.\ Rev.\ D, {\bf 51}: 1125 (1995);
   Erratum: [Phys.\ Rev.\ D, {\bf 55}: 5853 (1997)]
 % doi:10.1103/PhysRevD.55.5853, 10.1103/PhysRevD.51.1125
  [hep-ph/9407339]
  %%CITATION = doi:10.1103/PhysRevD.55.5853, 10.1103/PhysRevD.51.1125;%%
  %1866 citations counted in INSPIRE as of 05 Apr 2016
%\cite{Chen:2007vu}
\bibitem{Chen:2007vu}
  Y.~Chen {\it et al.} [CLQCD Collaboration],
  %``Radially excited states of 1P charmonium and X(3872),''
  hep-lat/0701021
  %%CITATION = HEP-LAT/0701021;%%
  %12 citations counted in INSPIRE as of 05 avril 2016
%\cite{Dudek:2006ej}
\bibitem{Dudek:2006ej}
  J.~J.~Dudek, R.~G.~Edwards and D.~G.~Richards,
  %``Radiative transitions in charmonium from lattice QCD,''
  Phys.\ Rev.\ D, {\bf 73}: 074507 (2006)
 % doi:10.1103/PhysRevD.73.074507
  [hep-ph/0601137]
  %%CITATION = doi:10.1103/PhysRevD.73.074507;%%
  %118 citations counted in INSPIRE as of 12 Apr 2016
%\cite{Dudek:2006ej}
%\cite{Gui:2012gx}
\bibitem{Gui:2012gx}
  L.~C.~Gui {\it et al.} [CLQCD Collaboration],
  %``Scalar Glueball in Radiative $J/\psi$ Decay on the Lattice,''
  Phys.\ Rev.\ Lett., {\bf 110}: 021601 (2013)
%  doi:10.1103/PhysRevLett.110.021601
  [arXiv:1206.0125 [hep-lat]]
  %%CITATION = doi:10.1103/PhysRevLett.110.021601;%%
  %29 citations counted in INSPIRE as of 12 Apr 2016
%\cite{Ablikim:2013wzq}
\bibitem{Ablikim:2013wzq}
  M.~Ablikim {\it et al.} [BESIII Collaboration],
  %``Observation of a Charged Charmoniumlike Structure $Z_c$(4020) and Search for the $Z_c$(3900) in $e^+e^- \to дл^+дл^-h_c$,''
  Phys.\ Rev.\ Lett.,  {\bf 111}: 242001 (2013)
%  doi:10.1103/PhysRevLett.111.242001
  [arXiv:1309.1896 [hep-ex]]
  %%CITATION = doi:10.1103/PhysRevLett.111.242001;%%
  %165 citations counted in INSPIRE as of 06 Apr 2016
%\cite{Dubynskiy:2008mq}

\bibitem{Dubynskiy:2008mq}
  S.~Dubynskiy and M.~B.~Voloshin,
  %``Hadro-Charmonium,''
  Phys.\ Lett.\ B, {\bf 666}: 344 (2008)
%  doi:10.1016/j.physletb.2008.07.086
  [arXiv:0803.2224 [hep-ph]]
  %%CITATION = doi:10.1016/j.physletb.2008.07.086;%%
  %99 citations counted in INSPIRE as of 06 Apr 2016

\end{thebibliography}
\end{document}